\documentclass[10pt,aps,prl,twocolumn,groupedaddress,showkeys]{revtex4-1}
\usepackage{amsmath}
\usepackage{graphicx}
\usepackage[colorlinks=true,linkcolor=blue,urlcolor=blue,citecolor=blue,anchorcolor=blue]{hyperref}
\DeclareGraphicsExtensions{.pdf,.eps,.png,.jpg,.mps}
\usepackage{epstopdf}
\usepackage{threeparttable}
\usepackage{color}

\begin{document}

\title{Graphene-Quantum Dots Hybrid Photodetectors with Low Dark-Current Readout}
\author{Domenico De Fazio$^1$}
\thanks{These authors contributed equally to this work}
\author{Burkay Uzlu$^{2,3}$}
\thanks{These authors contributed equally to this work}
\author{Iacopo Torre$^1$}
\author{Carles Monasterio$^1$}
\author{Shuchi Gupta$^1$}
\author{Tymofiy Khodkov$^1$}
\author{Yu Bi$^1$}
\author{Zhenxing Wang$^2$}
\author{Martin Otto$^2$}
\author{Max C. Lemme$^{2,3}$}
\author{Stijn Goossens$^1$}
\author{Daniel Neumaier$^{2,4}$}
\email[]{neumaier@amo.de}
\author{Frank H. L. Koppens$^1$}
\email[]{frank.koppens@icfo.eu}
\affiliation{$^1$ ICFO-Institut de Ciencies Fotoniques, The Barcelona Institute of Science and Technology, 08860 Castelldefels (Barcelona), Spain}
\affiliation{$^2$ Advanced Microelectronic Center Aachen (AMICA), AMO GmbH, 52074 Aachen, Germany}
\affiliation{$^3$ Chair of Electronic Devices, RWTH Aachen University, 52074 Aachen, Germany}
\affiliation{$^4$ Chair of Smart Sensor Systems, University of Wuppertal, 42119 Wuppertal, Germany}


\begin{abstract}
Graphene-based photodetectors have shown responsivities up to 10$^8$A/W and photoconductive gains up to 10$^{8}$ electrons per photon. These photodetectors rely on a highly absorbing layer in close proximity of graphene, which induces a shift of the graphene chemical potential upon absorption, hence modifying its channel resistance. However, due to the semi-metallic nature of graphene, the readout requires dark currents of hundreds of $\mu$A up to mA, leading to high power consumption needed for the device operation. Here we propose a novel approach for highly responsive graphene-based photodetectors with orders of magnitude lower dark current levels. A shift of the graphene chemical potential caused by light absorption in a layer of colloidal quantum dots, induces a variation of the current flowing across a metal-insulator-graphene diode structure. Owing to the low density of states of graphene near the neutrality point, the light-induced shift in chemical potential can be relatively large, dramatically changing the amount of current flowing across the insulating barrier, and giving rise to a novel type of gain mechanism. This readout requires dark currents of hundreds of nA up to few $\mu$A, orders of magnitude lower than other graphene-based photodetectors, while keeping responsivities of $\sim$70\,A/W in the infrared, almost two orders of magnitude higher compared to established germanium on silicon and indium gallium arsenide infrared photodetectors. This makes the device appealing for applications where high responsivity and low power consumption are required. 
\end{abstract}

\maketitle

Modern integrated photonics heavily relies on the complementary metal-oxide-semiconductor (CMOS) technology developed on silicon (Si)\cite{Reed2004}. The current photodetector scene is dominated by Si-based devices in the visible range (wavelength $\lambda\sim$400\,nm to 800\,nm)\cite{Reed2004}. However the $\sim$1.1\,eV bandgap of Si inhibits absorption at wavelengths $>$1.13\,$\mu$m\cite{Reed2004,Sze2006}. Si is therefore replaced by germanium (Ge) or III-V semiconductor ternary alloys such as indium gallium arsenide (InGaAs) in the near-infrared and specifically at the telecommunication-relevant wavelengths $\lambda\sim$1.55\,$\mu$m and $\lambda\sim$1.31\,$\mu$m\cite{Sze2006}. Ge grown epitaxially on Si is the current preferred choice in the CMOS industry\cite{KangNP2009} as the integration of III-V(s) in the CMOS fabrication line remains still a challenge\cite{Sze2006}. Furthermore, the complex process for the growth of Ge on Si\cite{WangS2011}, leaves room for more advantageous alternatives, also since back-end-of-line (BEOL) compatible integration schemes are required to exploit the full potential of 3-D integration.

Owing to the atomic thickness\cite{NovoS2004}, broadband absorption\cite{NairS2008} and ultra-high carrier mobility ($>$10$^5$\,cm$^2$V$^{-1}$s$^{-1}$) at room temperature\cite{MayoNL2011} of graphene, graphene-based optoelectronic devices feature amongst the ideal candidate photodetectors\cite{AkinN2019,NeumNM2019}. The production of graphene photodetectors\cite{GossNP2017,GoykNL2016} and modulators\cite{LiuN2011,SoriNP2018} has been proven to be compatible with BEOL-CMOS integration\cite{GossNP2017,GoykNL2016,LiuN2011,SoriNP2018}, promising fast, efficient and broadband device operation\cite{AkinN2019,NeumNM2019}. 

The performance of photodetectors can be quantified through several key figures of merit\cite{KoppNN2014}. The responsivity $R_{\rm ph}$, measured in A/W is the ratio between the collected photocurrent $I_{\rm ph}$ and the incident light power on the detector $P_{\rm in}$, where $I_{\rm ph}$ represents the difference taken in module between the current measured when shining light ($I_{\rm light}$) and in dark ($I_{\rm dark}$)\cite{KoppNN2014}, \textit{i.e.} $I_{\rm ph}$=$|I_{\rm light}$-$I_{\rm dark}|$. The gain ($G_{\rm ph}$) quantifies the ability of some photodetectors to extract more than one carrier upon absorption of one photon. Photodetectors possessing a gain mechanism typically yield higher $R_{\rm ph}$ but require, in turn, higher bias. The noise equivalent power (NEP) and noise equivalent irradiance (NEI) represent the lowest detectable input power $P_{in}$ and input power density $P_D$, respectively. Another important parameter of a photodetector is the time response $\tau_{res}$, which can be related to the frequency at which the device response becomes half of the response measured with unmodulated light.

\begin{figure*}
\centerline{\includegraphics[width=175mm]{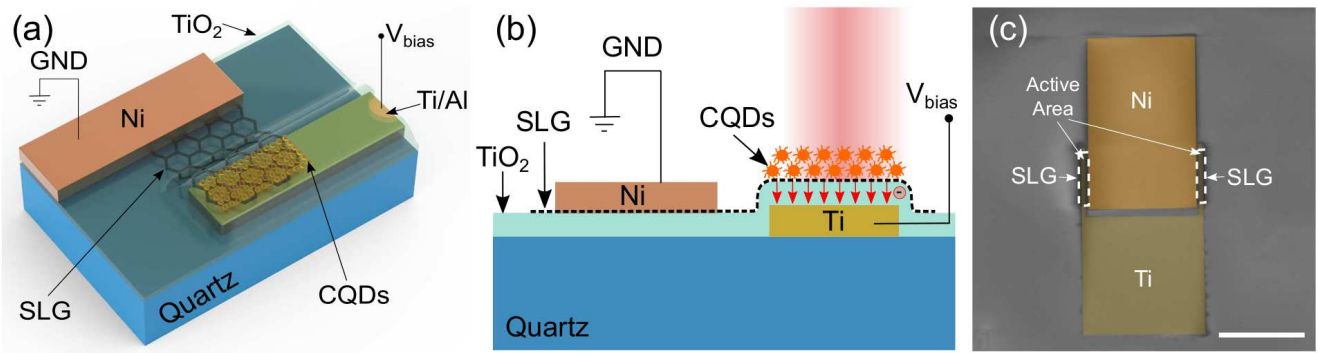}}
\caption{A sketch in perspective (a) and side view (b) of our metal-insulator-SLG/CQD photodetector. The biasing configuration is shown: a $V_{\rm bias}$ is applied to the Ti electrode, while the Ni/SLG electrode is set to ground (GND). (c) SEM image of one device highlighting the SLG area and the device active area, \textit{i.e.} the area that overlaps SLG with the Ti electrode, where also a layer of TiO$_2$ is present.}
\label{fig:Fig1}
\end{figure*}

Graphene field-effect-transistor-based (GFET) photodetectors have achieved outstanding performance with $R_{\rm ph}\sim 10^{8}$\,A/W\cite{KonsNN2012,ZhanSR2014} in the visible range and $R_{\rm ph}\sim 10^{6}$\,A/W in the near-infrared\cite{KonsNN2012}. In these readouts an absorber, such as a layer of colloidal quantum dots (CQDs)\cite{KonsNN2012} or the layered semiconductor MoS$_2$\cite{ZhanSR2014,DeFaACS2016}, are in direct contact with graphene in a GFET. Light impinging on the absorbing layer causes the formation of electron/hole pairs; while one carrier type is donated to the graphene channel, the other is trapped in the absorbing layer, causing a shift in the chemical potential of graphene, hence modifying its channel resistance\cite{KonsNN2012,ZhanSR2014,DeFaACS2016}. However, due to the semi-metallic nature of graphene\cite{NovoS2004}, a standard GFET readout scheme, relying on a graphene channel of area $A_{\rm ch}\sim$few tens of $\mu$m$^2$ in between two metal electrodes, can result in $I_{\rm dark}$ rapidly approaching the $\mu$A regime with a few mV bias voltage ($V_{\rm bias}$)\cite{XiaNN2009}. For comparison, germanium (Ge) on Si near-infrared photodiodes of similar footprint typically require few nA for operation\cite{DeRoOE2011,ColaIEEE2007}, at the expenses of responsivities $R_{\rm ph}<$1\,A/W \cite{DeRoOE2011,ColaIEEE2007}. One possible solution is to replace the semi-metallic graphene channel by a semiconducting channel of layered MoS$_2$ in Ref.\citenum{KufeAM2015}, yielding $R_{\rm ph}\sim 6\cdot 10^{5}$\,A/W, while keeping currents $<$1\,$\mu$A. However these devices showed a limited $\tau_{res}\sim$0.3\,s, which is not acceptable for many applications such as real-time imaging\cite{KoppNN2014}.

Here, we present a novel photodetector readout scheme with a novel type of gain mechanism while keeping the dark current in the nA range. This is achieved by sensitisation of a metal-insulator-graphene (MIG) diode structure\cite{ShayN2017,WangACSAEM2019} with CQDs as absorbing layer. The charge trapping upon light absorption in the CQDs layer induces a shift in the chemical potential of graphene which, in turn, generates a change in the current flowing across the oxide barrier. The shift in chemical potential can be relatively large due to the low density of states of graphene near the neutrality point\cite{CastRMP2009}. In turn, the shift in chemical potential of graphene can cause a relatively large change in the amount of current flowing across the barrier, due to the exponential dependency of the current on the diode barrier heigth, leading to a gain mechanism. This novel readout scheme leads to $R_{\rm ph}\sim 42$\,A/W in the visible and $\sim 70$\,A/W in the infrared while keeping a dark current in the nA range. The infrared responsivity is almost two orders of magnitude higher than Ge on Si infrared detectors\cite{DeRoOE2011,ColaIEEE2007}. Our $R_{\rm ph}$ is lower compared to some graphene-based\cite{KonsNN2012,ZhanSR2014,DeFaACS2016} and MoS$_2$-based photoconductor hybrids\cite{KufeAM2015}, however here we reduce $I_{\rm dark}$ down to the nA range, while still holding a gain mechanism together with a $\tau_{res}\sim$1.4$\cdot 10^{-4}$\,s. Another advantage of our device with respect to other graphene-based photodetectors is its weak dependence on the quality of graphene used in the fabrication. This is because its detection mechanism is based on out-of plane charge transport which only weakly relies on high in-plane charge-carrier mobilities. These features make the device an appealing candidate for low-power photo-detection applications where gain is required, also fulfilling the requirements needed for CMOS. 

\section{\label{Results}Results}

\begin{figure*}[htbp!]
\centerline{\includegraphics[width=175mm]{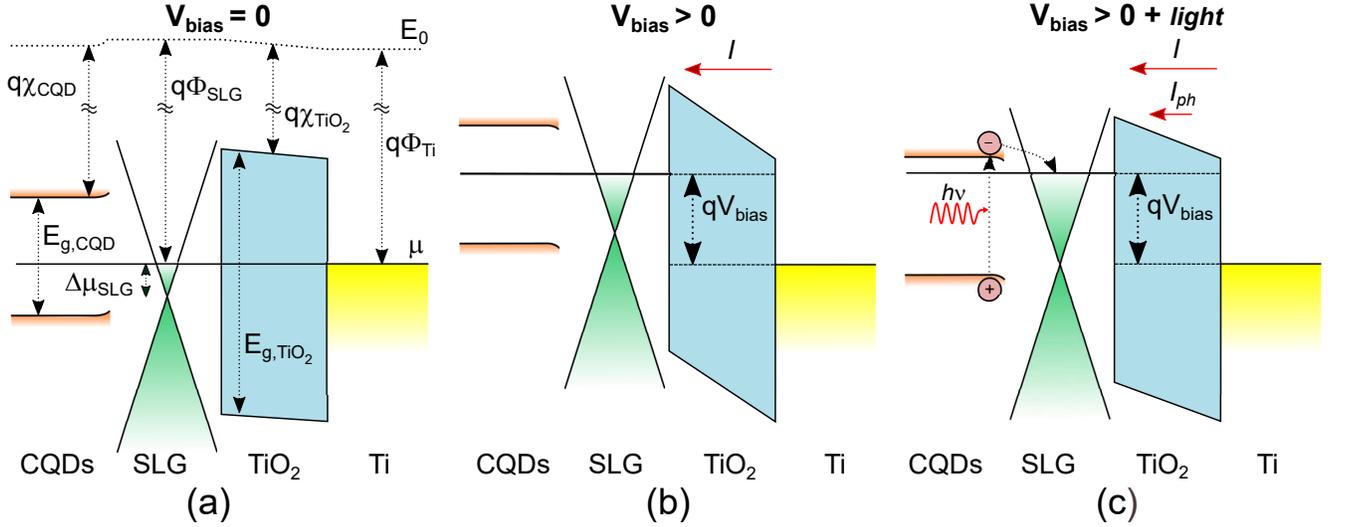}}
\caption{Qualitative representation of the structure band diagram (a) at $V_{\rm bias}=0$, (b) when applying a positive bias $V_{\rm bias}>0$ and (c) when adding light illumination to a $V_{\rm bias}>0$. In this figure $E_0$ is the vacuum level, $q\chi_{CQDs}$ and $q\chi_{TiO2}$ are CQDs and TiO$_2$ electron affinities, respectively, $q\Phi_{SLG}$  and $\Phi_{Ti}$ are SLG and Ti work functions, respectively, $E_{g,CQD}$ and $E_{g,TiO_2}$ are CQD and TiO$_2$ energy bandgaps, respectively, $\mu$ is the common chemical potential at $V_{\rm bias}=0$, $\Delta\mu_{SLG}$ is the shift in chemical potential of SLG due to doping and $h\nu$ is the photon energy.}
\label{fig:Fig2}
\end{figure*}

A schematic of the device is shown in Fig.\ref{fig:Fig1}(a), with a side view in Fig.\ref{fig:Fig1}(b). Conventional photo-lithography is used to pattern electrodes and form MIG diodes on a rigid quartz carrier substrate. First, a titanium/aluminium (Ti/Al) anode metal with thickness 10\,nm/20\,nm is sputter-deposited on the substrate, followed by a lift off process. From now on we will refer to this electrode as the Ti electrode for simplicity. A titanium oxide (TiO$_2$) layer of 6\,nm is then deposited \textit{via} an oxygen-plasma enhanced atomic layer deposition (ALD) process to form a dielectric barrier\cite{ShayN2017,WangACSAEM2019}. This is done at $T\sim$300$^{\circ}$C by using titanium tetrachloride (TiCl$_4$) as precursor for Ti. 

Commercially available chemical vapor deposited (CVD) single-layer-graphene (SLG), grown on 25\,$\mu$m thick copper (Cu)\cite{LiS2009} is transferred by means of a polymethyl-methacrylate (PMMA) assisted wet transfer method, using an aqueous iron(III)-chloride (FeCl$_3$) solution to etch the Cu substrate as described in Ref.\citenum{KimN2009}. PMMA is then dissolved in acetone. Next, graphene is patterned by means of reactive ion etching (RIE) with O$_2$ plasma using photolithography. Contact to graphene is provided by 25\,nm of sputter-deposited nickel (Ni) followed by a lift off process, which provides ohmic contact to graphene\cite{ShayAP2017}.

The result of this fabrication process is shown in a false colour scanning electron microscope (SEM) image in Fig.\ref{fig:Fig1}(c). We fabricated both detectors with a single graphene channel geometry and with two channels working in parallel, as per Fig.\ref{fig:Fig1}(c). Lead sulphide (PbS) CQDs of diameter $\sim$6\,nm have then been synthesized and spin-cast on the sample\cite{GossNP2017}, which results in a $\sim$100\,nm semiconducting solid-state CQD film with first excitonic peak around $\sim$1625\,nm (Supplementary Data 1). Details on the growth of CQDs and characterisation of the SLG and CQD layers are also reported in Supplementary Data 1. Samples are then wire-bonded on a chip carrier for measurements. 

The working principle of the device can be understood based on the following model. One can consider the device as composed of two parts, the metallic electrode on one side and the SLG plus the CQD layer on the other, weakly connected by the oxide barrier. In the absence of light and bias, the chemical potential is flat across the device, as shown in Fig.\ref{fig:Fig2}(a). When a certain voltage bias $V_{\rm bias}$ is applied, each of the two parts composing the structure remains in local equilibrium but a chemical potential difference $-qV_{\rm bias}$ ($-q$ being the electron charge) is established between them by the external voltage source {Fig.\ref{fig:Fig2}(b)}.
Taking into account the electrostatics of the structure and band alignments (see Supplementary Data 2) we obtain the equilibrium equation, which will also be valid under illumination:
\begin{equation}
\label{eq:Eq1}
V_{\rm bias} = 
(\Phi_{Ti} -\Phi_{SLG}^{0}) 
+ \frac{q (n_{SLG}- \Delta n)}{C} 
+ \frac{\Delta \mu_{SLG}}{q}.
\end{equation}
Here, $q\Phi_{Ti}$ is the titanium work function, $q\Phi_{SLG}^{0}$ the work function of undoped graphene, and $C$ the capacity per unit area of the junction.
The variable $n_{SLG}$ is the electronic density in the SLG layer, while $\Delta n = n_{Ti} + n_{SLG}$ the number of electrons per unit area transferred from the CQDs layer to the junction ($n_{Ti}$ being the electron density on the Ti surface). This charge transfer is present also in the dark due to charge redistribution between the different parts of the device and it depends on $V_{\rm bias}$; we checked the impact of this contribution with numerical simulations and decided to neglect it since it does not alter significantly the results. We also neglected, in our analysis, the presence of residual carrier concentration in graphene due to impurities, since it cannot be directly measured and it also has a minor impact on the results of our model.  However, in our photo-detection scheme, $\Delta n$ induced by illumination is a dominant factor. Finally, $\Delta\mu_{SLG} =  v_{\rm D} \text{sgn}(n_{SLG})\sqrt{\pi |n_{SLG}|}$ is the shift in chemical potential due to the change in charge carrier concentration\cite{DasNN2008}, $v_{\rm D} \approx 10^6\,{\rm  m/s}$ being the Dirac velocity in graphene\cite{GeimNM2007}.
The first term in Eq.(\ref{eq:Eq1}) describes the charge transfer due to the difference in work function between SLG and Ti; the second is the electrostatic potential difference due to the redistribution of charges, while the last term accounts for the quantum capacitance of SLG\cite{XiaNN2009-2}, or in simpler words, for the change of the chemical potential due to a variation of the electronic density. The latter effect is negligible in three dimensional metallic systems due to their large density of states that can easily accommodate additional electrons without significant changes in the work function, but plays an important role in graphene because of its vanishing density of state close to the charge neutrality point\cite{XiaNN2009-2}. Solving Eq.\ref{eq:Eq1} allows us to draw the band diagram of Fig.\ref{fig:Fig2}(b).

\begin{figure*}[ht!]
\centerline{\includegraphics[width=175mm]{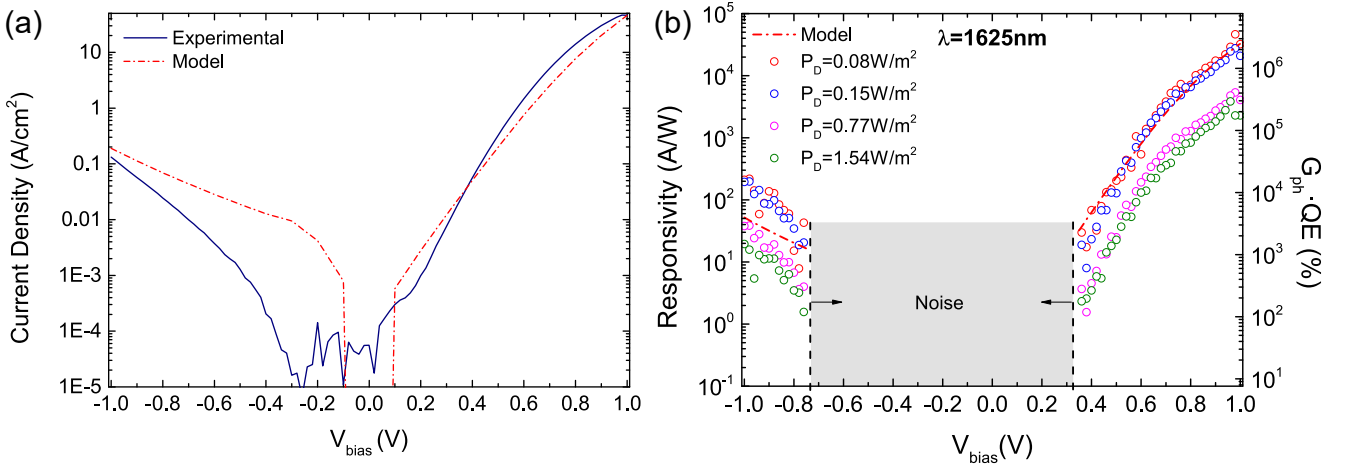}}
\caption{(a) Current density-bias voltage device characteristics in dark. The red dash-dot line is the trend of current density predicted by our theoretical model, while the solid dark-blue line is measured experimentally. (b) Device responsivity $R_{\rm ph}$ (left axis)/external quantum efficiency (right axis) upon light illumination when using a $\lambda$=1625\,nm laser, measured at increasing power irradiances ($P_D$) (coloured circles). The red dash-dot line is the trend of $R_{\rm ph}$ predicted by our theoretical model. Data within -0.74\,V$<V_{\rm bias}<$0.34\,V has been hidden as the light response falls within the noise level of the used instrumentation.}
\label{fig:Fig3}
\end{figure*}

Electrons can travel through the barrier due to its finite transmittivity. The total current depends on the applied bias and the height of the barrier. These two factors determine the asymmetric dark current density-voltage characteristics. Using a simple model\cite{ShayN2017,BritS2012,AliaAPL2016}, as discussed in Supplementary Data 2, we calculate the flow of current in the device, obtaining the dash-dot red line in Fig.\ref{fig:Fig3}(a). Our model takes into account energy-conserving jumps of electrons between the two sides of the oxide barrier. 
The number of jumps per unit time at a given energy is controlled by the number of possible starting and final states at that energy, quantified by the densities of states, by the occupation of the initial and final states, determined by the Fermi distribution, and by an energy dependent probability of transmission across the barrier\cite{Griffiths2018}. The total current is then proportional to the sum over energies of the number of jumps per unit time at each energy.

A fraction of the light impinging on the device is absorbed and generates electron-hole pairs in the CQDs layer. A smaller fraction of these photo-generated carriers are then transferred to the SLG layer. The external quantum efficiency ${\rm QE}$ measures this fraction of electrons absorbed and transferred to SLG, triggered by absorption of one photon.

The amount of charge $\Delta n(V_{\rm bias},P_D)$ added to the diode area $A_{\rm ph}$ due to light illumination at small $P_D$ is:
\begin{equation}
\label{eq:Eq2}
\Delta n(V_{\rm bias},P_D)\approx \Delta n(V_{\rm bias},0) + \tau_{\rm tr} {\rm QE}\frac{P_D}{E_{\rm ph}}.
\end{equation}
Here $E_{\rm ph}$ is the energy of a single incident photon, which can be calculated as $E_{\rm ph}=hc/\lambda$, with $h$ being the Planck constant, $c$ being the speed of light in vacuum, and $\tau_{\rm tr}$ is the average time a photo-generated electron spends in the device. It can be noted that $P_D/E_{\rm ph}$ is the number of photons per unit of area and time impinging on the device. We assume that $\tau_{\rm tr}$ does not depend on $V_{\rm bias}$.

As a result, the chemical potential of graphene $\Delta\mu_{SLG}$ grows because of the presence of these additional electrons according to Eq.\ref{eq:Eq1} (see also Fig.\ref{fig:Fig2}(c)), while the bias is kept constant. This produces a lowering of the barrier on the SLG side that enhances the current with respect to the value in absence of illumination. This extra current, \textit{i.e.} $I_{\rm ph}$, represents the light signal of our photodetector. The expression for QE in relation to $I_{\rm ph}$ and $R_{\rm ph}$ is then\cite{KoppNN2014}:

\begin{equation}
\label{eq:Eq3}
{\rm QE}=\frac{I_{\rm ph}/q}{P_{\rm in}/E_{\rm ph}}\cdot\frac{1}{G_{\rm ph}}=\frac{R_{\rm ph}}{qE_{\rm ph}}\cdot\frac{1}{G_{\rm ph}}.
\end{equation}

Here $G_{\rm ph}$ is the gain and it accounts for the fact that one single photon can produce a distortion of the barrier that affects the motion of many other electrons. This results in multiple charge carrier detections upon absorption of a single photon, leading to a gain mechanism. Ge and InGaAs photodiodes do not possess a gain mechanism ($G_{\rm ph}$=1), then the overall charge extraction efficiency QE$\cdot G_{\rm ph}$ is limited up to 100\%\cite{MichNP2010,KaniOER2004}.

Making use of Eq.\ref{eq:Eq1}-\ref{eq:Eq2} and the conduction model detailed in Supplementary Data 2 it is possible to calculate the responsivity of the photodetector in the regime of small powers. The combination of the two equations yields:
\begin{equation}
\label{eq:Eq4}
R_{\rm ph}(V_{\rm bias},P_D \to 0) =\frac{q\tau_{\rm tr} {\rm QE}}{h\nu R_{\rm tr} C_{\rm tot}},
\end{equation}
where $C_{\rm tot}= A_{\rm ph}(C +C_{\rm diff})$ is the total capacity, $C_{\rm diff} \equiv e^2 \partial n_{SLG}/\partial\Delta \mu_{SLG}$ being the differential quantum capacitance of SLG, and $R_{\rm tr}$ is the transfer resistance given by $R_{\rm tr} \equiv \left(\partial qI/\partial \Delta\mu_{SLG}\right)^{-1}$. 
Here the derivative should be taken at zero power and constant $V_{\rm bias}$. 
The transfer resistance can also be estimated using the differential resistance in dark, {\it i.e.} $R_{\rm tr} \approx \left(\partial I_{\rm dark}/\partial V_{\rm bias}\right)^{-1}$.
These two numbers can be calculated within our model and the final result gives a theoretical trend for $R_{\rm ph}$ shown in Fig.\ref{fig:Fig3}(b) as a red dash-dot line.

We now discuss experimental results, measuring the response of the device in dark and upon light illumination. The bias $V_{\rm bias}$ is supplied to the sample by a data acquisition system (DAQ) board, while the drain electrode is connected to an Ithaco Current Preamplifier, and then back to the DAQ to read the output current $I$. All measurements are taken in vacuum. The dark-blue curve in Fig.\ref{fig:Fig3}(a) represent the current density-voltage device characteristics when a $V_{\rm bias}$ voltage is applied to the Ti electrode, while the Ni/graphene contact is grounded. The current density $J$ is calculated as $I/A_{\rm ph}$, with $A_{\rm ph}$ being the area of the barrier region, also corresponding to the photoactive area. The electrical behaviour of the device follows the expected asymmetric diode-like behaviour of our theoretical model, while using only one fit parameter (an overall multiplicative constant).

For the optoelectronic measurements of the device, we use laser diodes of wavelengths $\lambda$=685\,nm and $\lambda$=1625\,nm. We modulate the light impinging on the device \textit{via} a sinusoidal 30\,Hz signal from an Agilent Function Generator. We then record a current signal of 10 seconds ($\Delta t$) using 10\,kHz sampling frequency ($F_S$) in order to measure the signal power spectral density $S_I$. Specifically, $S_I$, which is a function of frequency $f$ and it is measured in A$^2$/Hz units, can be extracted by performing the fast Fourier transform (FFT) of the recorded signal\cite{Brown1992}:

\begin{equation}
\label{eq:Eq5}
S_I(f)=\frac{1}{\Delta t\cdot F_S}\cdot|\text{FFT}|^2.
\end{equation}

A peak appearing at 30\,Hz in the $S_I$ trend plotted as a function of $f$ ($S_I|_{30Hz}$) is thus the fingerprint of the device being able to detect the light signal. The responsivity $R_{\rm ph}$ of the device can then also be calculated as:

\begin{equation}
\label{eq:Eq6}
R_{\rm ph}=\frac{2\sqrt{2\cdot S_I|_{30Hz}}}{P_D\cdot A_{\rm ph}}=\frac{I_{\rm ph}}{P_D\cdot A_{\rm ph}}=\frac{I_{\rm ph}}{P_{\rm in}},
\end{equation}

where $P_D$ is the power density value at which we measure $S_I|_{30Hz}$ and the device photoactive area $A_{\rm ph}$ is $\sim$210\,$\mu$m$^2$. 

Shining light results in an increase of the current flowing across the barrier, as anticipated in Fig.\ref{fig:Fig2}(c), at sufficiently positive (or negative) bias. Fig.\ref{fig:Fig3}(b) plots the device $R_{\rm ph}$ as a function of $V_{\rm bias}$ at several light power densities $P_D$ and when shining infrared light. A detectable photo-induced signal is present at $V_{\rm bias}\lesssim -0.74$V and $V_{\rm bias}\gtrsim 0.34$\,V. At -0.74\,V$\lesssim V_{\rm bias}\lesssim 0.34$\,V the signal is too low and disappears in the noise, due to the use of bulk readout instruments. $R_{\rm ph}$ is higher at lower $P_D$ and it reaches values as high as $\sim 3\cdot 10^{4}$\,A/W at 1\,V and $P_D=$0.08\,W/m$^2$. However, optimal $R_{\rm ph}$ at operating $ V_{\rm bias}\sim$0.5-0.8\,V (low dark current regime) will be determined later on with power-dependent measurements. The theoretical trend of $R_{\rm ph}$ is in reasonable agreement with experimental data, even though only one fit parameter (an overall multiplicative constant) was used. An $R_{\rm ph}$ much higher than 1\,A/W suggests the presence of a gain mechanism, as for photoconductors, meaning that multiple electrons are collected upon absorption of one photon\cite{KoppNN2014}. 

$QE\cdot G_{\rm ph}$, shown on the right axis of Fig.\ref{fig:Fig3}(b), reaches values up to $\sim 10^{4}\%$ at the operating voltage 0.5\,V and values above $10^{6}\%$ at 1\,V. The magnitude of the two components ($QE$ and $G_{\rm ph}$) is difficult to unravel in our device, as these depend on unknown parameters such as the charge trapping time and the charge-transfer efficiency at the interface between CQD and SLG.

\begin{figure*}[ht!]
\centerline{\includegraphics[width=175mm]{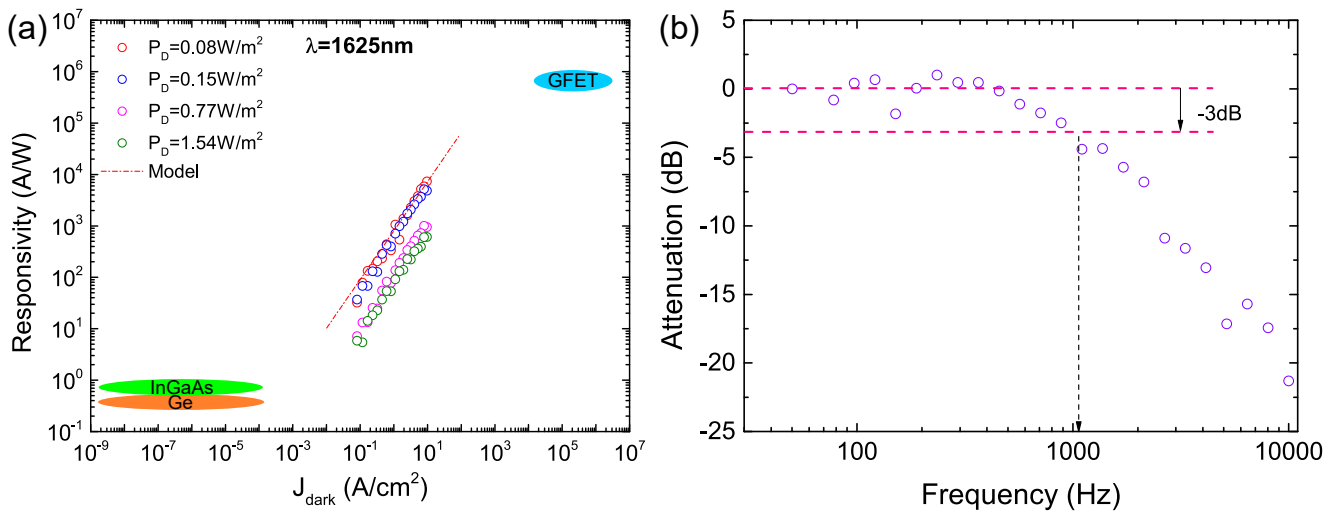}}
\caption{(a) Responsivity $R_{\rm ph}$ as a function of current density $J_{\rm dark}$ flowing in the device without illumination, plotted at four different power densities $P_D$. The red dash-dotted line is the result of our theoretical model. Coloured areas on the bottom-left and top-right of the graph highlight the regions of $R_{\rm ph}$ versus $J_{\rm dark}$ where modern InGaAs, Ge and lateral GFET-based photodetectors stand.
(b) Bode diagram of the photoresponse attenuation (dB units) as a function of frequency, with emphasis on the 3\,dB cutoff.}
\label{fig:Fig4}
\end{figure*}

To further compare with Ge and InGaAs technologies, we plot in Fig.\ref{fig:Fig4}(a) $R_{\rm ph}$ as a function of current density $J$ in dark ($J_{\rm dark}$) at 1625\,nm wavelength. The latter is the current density flowing in the device in dark at the $V_{\rm bias}$ necessary to achieve that $R_{\rm ph}$ upon illumination. Here we focused on positive $V_{\rm bias}$ as a stronger light response can be achieved compared to the negative branch. The $J_{\rm dark}$ flowing in our devices around 0.1-1\,A/cm$^{2}$ (\textit{i.e.} 0.5-0.6\,V $V_{\rm bias}$) are at least two orders of magnitude higher than Ge and InGaAs photodiodes, but they also allow us to deliver about two orders of magnitude improvement in $R_{\rm ph}$. Other CQD-SLG hybrid photodetectors relying on lateral transport in a GFET configuration can yield $R_{\rm ph}$ up to 10$^6$\,A/W in the infrared, but at the expenses of $J_{\rm dark}\sim$10$^6$A/cm$^{2}$\cite{KonsNN2012}.

We then characterise the device time response monitoring the height of $S_I$ while varying the frequency of the sinusoidal signal modulating the laser light. We use a $V_{\rm bias}=0.6$\,V and 1\,W/m$^2$ power density at $\lambda$=1625\,nm. Fig.\ref{fig:Fig4}(b) plots the signal attenuation, defined as $10log_{10}(S_{I}|_{f_m}/S_{I}|_{30Hz})$, where $S_{I}|_{f_m}$ is the $S_I$ measured at frequency $f_m$ while modulating light with a sinusoid of frequency $f_m$. The signal halves (or decreases by 3\,dB) at around 1.1\,kHz, which is known as the cut-off frequency $f_{\text{cut-off}}$. This determines the device time response $\tau_{res}=1/(2\pi f_{\text{cut-off}})=1.4\cdot 10^{-4}$\,s, \textit{i.e.} 0.14\,ms, likely limited by the lifetime of photo-generated electrons. This is about three orders of magnitude faster than high-responsivity photodetectors relying on MoS$_2$\cite{KufeAM2015}.

\begin{figure*}[ht!]
\centerline{\includegraphics[width=180mm]{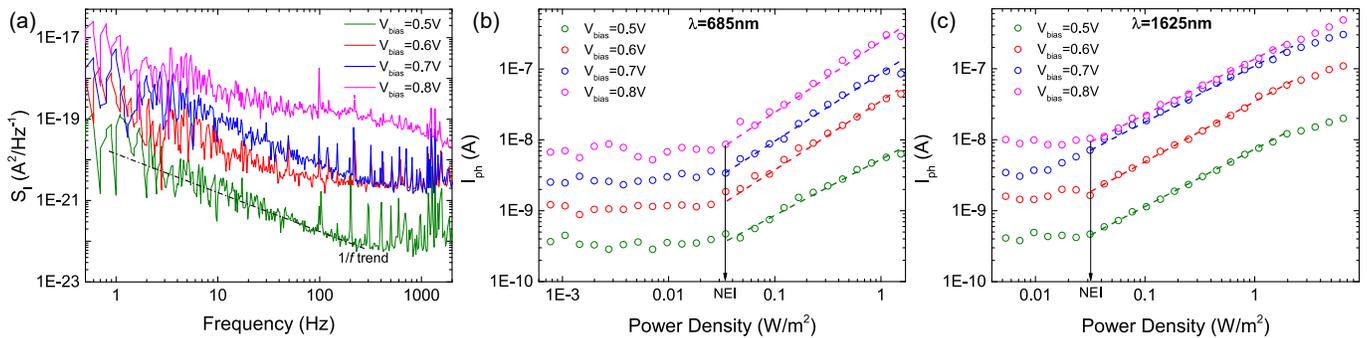}}
\caption{(a) Signal power spectral density measured as a function of frequency, in the range 0.5-1000\,Hz, at $V_{\rm bias}$ from 0.5\,V to 0.8\,V, \textit{i.e.} when the diode is ON. (b) Intensity of the signal power spectral density at 30\,Hz as a function of the laser power density irradiating the sample, measured at a wavelength $\lambda$=685\,nm and (c) $\lambda$=1625\,nm. The signal is flat and mixes with instrumental noise at power densities $<\sim0.03$\,W/m$^{2}$, then grows almost linearly at power densities $>\sim0.03$\,W/m$^{2}$. 0.03\,W/m$^{2}$ is thus taken as the device NEI.}
\label{fig:Fig5}
\end{figure*}

The importance of noise is crucial in photodetectors to determine figures of merit such as NEI and NEP\cite{KoppNN2014}. The predominant sources of noise in graphene devices are a combination of local fluctuations in the charge carrier density\cite{BalaNN2013}, local fluctuations in the carrier mobility\cite{ZhanACSN2011} and contact resistance\cite{KarnNC2016}. In dark, we observe at each bias a $1/f$ trend of $S_I$, which corresponds to Flicker noise, and it is the typical type of noise observed in graphene-based devices\cite{MavrN2018}, as well as in other 3-D material systems\cite{HoogRPP1981,HoogIEEE1994} such as metal-insulator-semiconductor transistors\cite{ZhigJCTE2007} and photodetectors\cite{MonrIEEE2000}. Specifically, Ref.\citenum{BalaNN2013} attributed the origin of Flicker noise in graphene to the random trapping/de-trapping of charge carriers at the interface with the oxide supporting the graphene layer, which ultimately affects transport across the material. In 3-D metal-insulator-semiconductor structures, noise is also originated from the charge trapping in the insulating layer, influencing electronic transport\cite{ZhigJCTE2007,MonrIEEE2000}. These mechanisms explain the overall $1/f$ trend observed in Fig.\ref{fig:Fig5}(a). At frequencies $>$40\,Hz, we start detecting the 50\,Hz mains hum and its harmonics, which explains the peaks in the figure. A larger bias causes $S_I$, to rise accordingly, which is typical of Flicker noise in electronic devices where $S_I$ is proportional to the average squared direct current $I^2$ measured at the respective bias\cite{MonrIEEE2000,MavrN2018}. This bias-dependent $1/f$ behaviour of $S_I$ leads to a trade-off when choosing the optimum operating point in our photodetectors, as higher biases can eventually yield a better photo-signal, but also an increase in the electronic noise, other than an increase in the power consumption.  

In order to benchmark the device noise we calculate the parameter $\beta$, defined as\cite{StolAPL2015}:

\begin{equation}
\label{eq:Eq7}
\beta=\frac{S_I}{I^2}\cdot A_{\rm ch} \cdot f,
\end{equation}

where $A_{\rm ch}$ is the graphene channel area and $f$ is the frequency at which $\beta$ is calculated, which is typically set to 1\,Hz. Furthermore, here we use $A_{\rm ch}$ rather than $A_{\rm ph}$ as also the area off the barrier/junction region may originate $1/f$ noise. The unit of $\beta$ is [$\mu$m$^{2}$]. As $S_I$ is proportional to $I^2$ in our $1/f$-dominated devices, $S_I/I^2$ is independent of bias. We can thus calculate a $\beta$ of  $\sim$4$\cdot$10$^{-5}$\,$\mu$m$^{2}$ at 1Hz, when having a $A_{\rm ch}\sim$595\,$\mu$m$^2$. This number is consistent with that obtained in other graphene devices on SiO$_2$/Si substrates\cite{BalaNN2013,StolAPL2015} and could be reduced down to $\beta$ in the order of 10$^{-9}$\,$\mu$m$^{2}$ replacing TiO$_2$ with hexagonal boron nitride (hBN)\cite{StolAPL2015}.

From these noise curves, we can estimate the device NEI by performing a power-dependence study at $V_{\rm bias}$ ranges of 0.5-0.8\,V for excitation of visible and near-infrared light. As seen in Fig.\ref{fig:Fig5}(b), $I_{\rm ph}=2\sqrt{2\cdot S_I|_{30Hz}}$ is flat at small power densities $P_D<$0.03\,W/m$^{2}$, as the device photoresponse is masked by the noise. Regardless of the bias that is used, $I_{\rm ph}$ then grows linearly in the range $\sim$0.03-1\,W/m$^{2}$.  We can thus adopt 0.03\,W/m$^{2}$ as the device NEI, independent on the applied bias voltage. The NEI remains unchanged in the infrared as shown in Fig.\ref{fig:Fig5}(c). At power densities $>$1\,W/m$^{2}$, the slope of $I_{\rm ph}$ changes, suggesting saturation of the CQD absorption\cite{KonsNN2012}. The device NEP=NEI$\cdot A_{\rm ch}$ is $\sim$1.8$\cdot 10^{-11}$\,W. Using $I_{\rm ph}$ in Fig.\ref{fig:Fig5}(b) and (c) we can extract $R_{\rm ph}$ through Eq.\ref{eq:Eq6} at optimal $P_D$, \textit{i.e.} at $P_D$ close to the NEI, both for visibile and infrared light. We obtain a $R_{\rm ph}\sim$42\,A/W in the visible and $R_{\rm ph}\sim$70\,A/W in the infrared at $V_{\rm bias}=0.5$\,V, corresponding to an $I_{\rm dark}\sim700$\,nA ($J_{\rm dark}\sim0.3$\,A/cm$^{2}$). The infrared responsivity is about two orders of magnitude higher than that obtained with Ge on Si devices\cite{DeRoOE2011,ColaIEEE2007,MichNP2010} and InGaAs photodiodes\cite{KaniOER2004}, while the novel readout presented keeps $I_{\rm dark}$ in the nA regime. 

\section{\label{Disc}Discussion}

These numbers confirm the potential of our novel device concept as a compelling candidate for efficient photodetectors for infrared wavelengths. The readout method can pave the way for cheap, low-power consumption graphene-based image sensors. This would create an alternative to Ge detectors in CMOS technologies or InGaAs cameras, whose prohibitive price is currently hindering the diffusion of such technology on a large scale. Optimisation of the device geometry, interfaces and oxide material/thickness could further boost the performance of the photodetector in terms of noise suppression, which would allow light-harvesting at lower optical power densities. Furthermore, the readout concept utilised in this work and its relative gain mechanism could be attained with alternative sensitising layers other than the CQDs, such as perovskites or biomolecules.

\section{\label{Ackn}Acknowledgements}
We thank Bruno Riccò for useful discussions and Marc Montagut for graphical support. F.H.L.K. acknowledges financial support from the Government of Catalonia trough the SGR grant, and from the Spanish Ministry of Economy and Competitiveness, through the “Severo Ochoa” Programme for Centres of Excellence in R\&D (SEV-2015-0522), support by Fundacio Cellex Barcelona, Generalitat de Catalunya through the CERCA program,  and the Mineco grants Plan Nacional (FIS2016-81044-P) and the Agency for Management of University and Research Grants (AGAUR) 2017 SGR 1656.  Furthermore, the research leading to these results has received funding from the European Union Seventh Framework Programme under grant agreements no. 785219 (Core2) and no. 881603 (Core3) Graphene Flagship. This work was supported by the ERC TOPONANOP under grant agreement no. 726001. This work was also financially supported by the German Science Foundation (DFG) within the priority program FFlexCom Project “GLECS” (contract no. NE1633/3). I.T. acknowledges funding from the Spanish Ministry of Science, Innovation and Universities (MCIU) and State Research Agency (AEI) via the Juan de la Cierva fellowship no. FJC2018-037098-I.

\pagebreak
\widetext
\begin{center}
\textbf{\large Supplemental Materials: Low Dark-Current Readout for Graphene-Quantum Dots Hybrid Photodetectors}
\end{center}

\setcounter{equation}{0}
\setcounter{figure}{0}
\setcounter{table}{0}
\setcounter{page}{1}
\makeatletter
\renewcommand{\theequation}{S\arabic{equation}}
\renewcommand{\thefigure}{S\arabic{figure}}
\renewcommand{\bibnumfmt}[1]{[S#1]}
\renewcommand{\citenumfont}[1]{S#1}

\begin{NoHyper}

\section{Supplementary Data 1: Growth and Characterisation}
\label{sec:S1}

\subsection{CQD Growth}

The growth of CQDs is performed by injecting the sulfur precursor (octadecene-diluted bis(trimethylsilyl)sulfide [(TMS)$_2$S]) in a three neck round bottom flask, using a syringe pump\cite{LeeS2016}. The flask also contains a lead (Pb)-oleate solution, which acts as the lead precursor\cite{LeeS2016}. This is prepared by mixing lead oxide (PbO), oleic acid and octadecene at $\sim90^{\circ}$C overnight under vacuum. The solution is then heated at $\sim100^{\circ}$C to finalise the CQD preparation\cite{LeeS2016}. Furthermore, as-synthesized CQDs were cadmium chloride (CdCl$_2$)-treated for passivation\cite{MihiAM2014,IpNN2012}. The PbS CQDs were washed with ethanol three times before making devices. A layer-by-layer method was then used to spin cast the CQDs on the sample to build a layer of $\sim$100\,nm\cite{SuppKonsNN2012}. A ligand exchange process was performed while spin-casting, in order to replace the long oleate ligands, resulting from CQDs growth, with shorter bidentate ligands of ethanedithiol (EDT)\cite{SuppKonsNN2012}. This turns the CQD-layer in a semiconducting solid state film\cite{SuppKonsNN2012}. 

\subsection{Characterization of SLG and CQDs}

The quality of graphene after transfer on the quartz substrate is monitored by performing Raman spectroscopy at an excitation wavelength of $\lambda\sim532$\,nm. We used a Renishaw InVia spectrometer equipped with 100$\times$ objective lens and kept laser power $<$1\,mW to avoid any damage or heating effect. Three representative spectra measured in proximity of the active area are shown in Fig.\ref{fig:FigS1}(a). The most prominent feature of graphene is the 2D peak\cite{FerrPRL2006}. Here the 2D peak at $\sim$2682\,cm$^{-1}$ is a single Lorentzian with average full-width-at-half maximum (FWHM) of $\sim$33\,cm$^{-1}$, a proof of single layer graphene\cite{FerrPRL2006,FerrNN2013}. The D peak around $\sim$1350\,cm$^{-1}$ is undetectable indicating negligible defects and a good transfer method\cite{CancNL2011}.

\begin{figure}[ht!]
\centerline{\includegraphics[width=120mm]{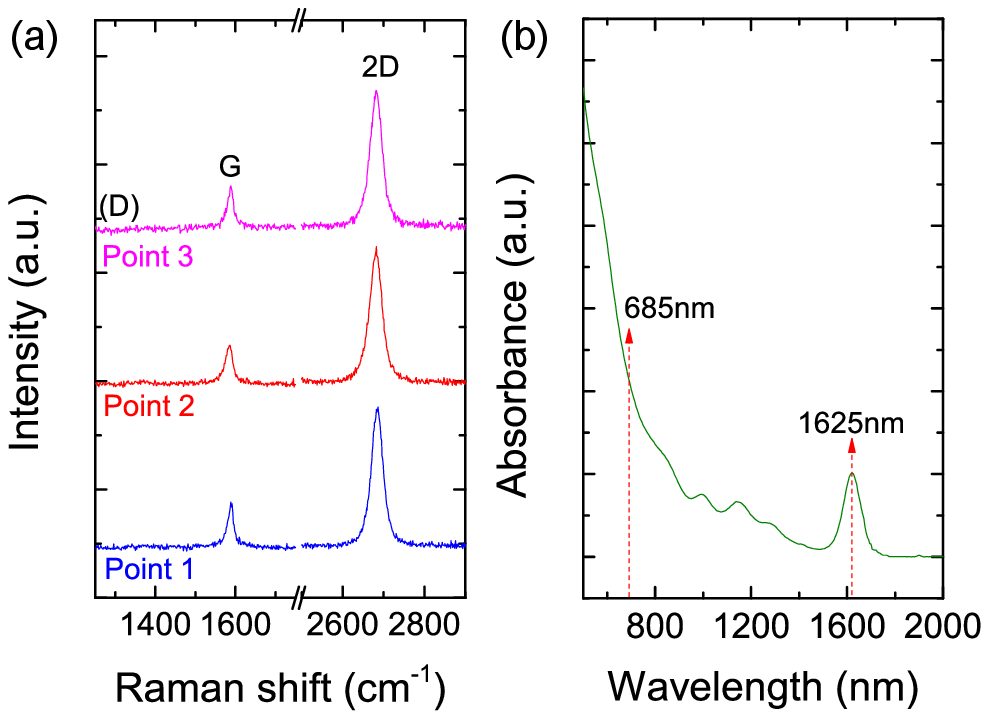}}
\caption{(a) Three representative Raman spectra of graphene wet-transferred on quartz + a $\sim$6nm thin TiO$_2$ layer, measured in random points of the graphene channel at $\lambda\sim$532\,nm. (b) UV-Vis-NIR absorbance spectrum of the Cd-Cl$_2$-treated CQD solution. The two wavelengths utilised for optoelectronic measurements are highlighed with dashed arrows.}
\label{fig:FigS1}
\end{figure}

Fig.\ref{fig:FigS1}(b) shows the absorbance spectrum of the fabricated CQDs solution, measured with a UV-Vis-NIR spectrometer. The excitonic peak at $\sim$1625\,nm is in correspondence with one of the two wavelengths utilized during the optoelectronic measurements, as expected. The other wavelength utilized in this work (685\,nm) is also highlighted in the graph.

\section{Supplementary Data 2: Theory}
\label{sec:S2}

\subsection{Band Diagram}

Here we describe how the band diagram of our device is calculated.
The band diagram requires the knowledge of the electrostatic potential and other material-related parameters. The latter are summarized in Fig.\ref{fig:FigS2} and Tab.\ref{tab:Tab1}.
\begin{figure}[ht!]
\centerline{\includegraphics[width=75mm]{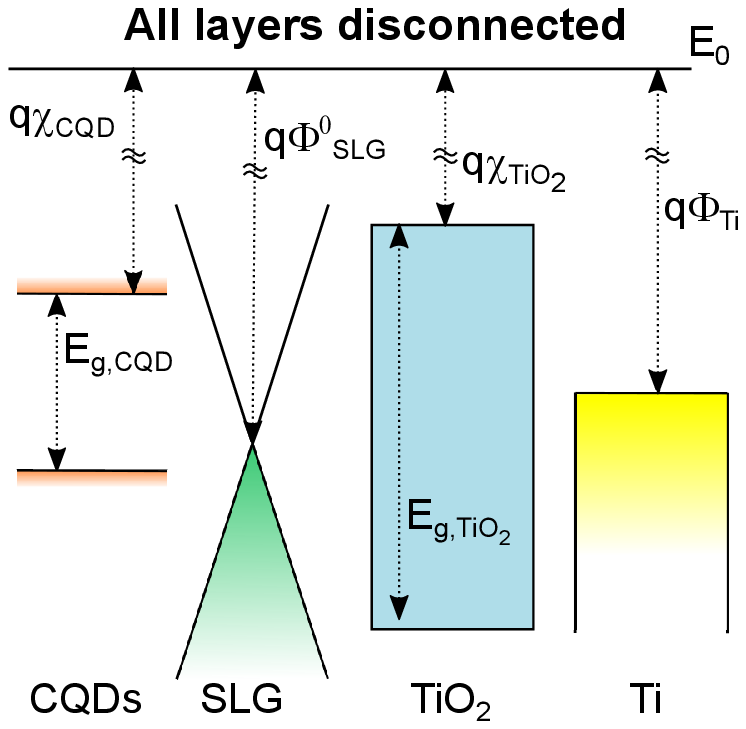}}
\caption{Band diagram of the different parts of the device with all layers forming the heterostructure disconnected}
\label{fig:FigS2}
\end{figure}
\begin{table}[ht!]
\begin{tabular}{lccc}
Parameter & Value & Unit & Reference\\
\hline
\hline
$q\chi_{CQD}$ &$3.9$ & eV & Ref.\citenum{BiN2019}\\
$E_{g,CQD}$ & $ 0.7$ & eV & Ref.\citenum{BiN2019}\\
$q\Phi_{SLG}^0$ & $4.6$ & eV & Refs.\citenum{YuNL2009,SeoJAP2014}\\
$q\chi_{TiO_2}^0$ & $4.0$ & eV & Ref.\citenum{BoroPCCP2011}\\
$E_{g,TiO_2}$ & $3.0$ & eV & Ref.\citenum{BoroPCCP2011}\\
$\epsilon_{TiO_2}$ & $15$ & -& Ref.\citenum{SuppWangACSAEM2019}\\
$m^*_{TiO_2}$ & $1$ & $m_{\rm e}$ & -\\ 
$q\Phi_{Ti}$ & $4.3$ & eV & Ref.\citenum{Weast1988}\\
$k_{\rm B}  T$ & $0.026$ & eV & -\\
\hline      
\end{tabular}
\caption{\label{tab:Tab1} Parameters used in device simulation}
\end{table}
Furthermore, the thickness of the CQD layer and the TiO$_2$ are measured to be $\sim$100\,nm and $\sim$6\,nm, respectively. 

In order to calculate the {\it electrostatic} potential in the structure we use Gauss's theorem.
The electric field $\textbf{E}$ is directed along the direction $z$ perpendicular to the layers forming the structure. We fix the origin of $z$ at the surface of the Ti electrode and we get:
\begin{equation}
\label{eq:EqS1}
\epsilon(z) E_z(z) - \epsilon(0^-) E_z(0^-) = -\frac{q}{\epsilon_0}\int_{0^-}^z d z' n(z') ,
\end{equation}
where $\epsilon(z)$ is the relative dielectric constant at the position $z$, and $n(z)$ is the electronic density.
The second term in Eq.\ref{eq:EqS1} is zero because the electric field vanishes in the metal. 
We can then write the electronic density as:
\begin{equation}
\label{eq:EqS2}
\begin{split}
n(z)  & = n_{Ti}\delta(z) + n_{SLG} \delta(z- t_{TiO_2}) \\
& +n_{CQD}(z) \theta(z- t_{TiO_2})\theta(t_{TiO_2}+t_{CQD}-z) ,
\end{split}
\end{equation}
where $\delta(z)$ and $\theta(z)$ are the Dirac delta and Heaviside step function respectively. 
In writing Eq.\ref{eq:EqS2} we took into account the fact that no charge is accumulated in the oxide barrier, while surface charges $n_{Ti}$ and $n_{SLG}$ accumulate at the surface of Ti and in the SLG layer.
Combining Eqs.\ref{eq:EqS1}-\ref{eq:EqS2} yields:
\begin{equation}
\label{eq:EqS3}
E_z(z) = 
\begin{cases}
0 \quad\mbox{if } z<0 \\
-\dfrac{qn_{Ti}}{\epsilon_0 \epsilon_{TiO_2}} \quad \mbox{if }0<z<t_{TiO_2} \\
-\dfrac{q\left[n_{Ti}+n_{SLG} + \int_{t_{TiO_2}}^zn_{CQD}(z')dz'\right]}{\epsilon_0 \epsilon_{CQD}} \quad\mbox{if } t_{TiO_2}<z< t_{TiO_2}+ t_{CQD}\\
0 \quad\mbox{if } t_{TiO_2}+ t_{CQD} <z .\\
\end{cases}
\end{equation}
Note that in order for the electric field to vanish outside the device the total charge should be zero, \textit{i.e.} $n_{Ti} + n_{SLG} = \Delta n$, where:
\begin{equation}
\label{eq:EqS4}
\Delta n = -\int_{t_{TiO_2}}^{t_{TiO_2}+t_{QD}} dz' n_{QD}(z'),
\end{equation}
is the total number of electrons per unit area transferred from the CQD layer to the rest of the device. The electrostatic potential $\phi(z)$ can be calculated by integrating:
\begin{equation}
\label{eq:EqS5}
\phi(z) = -\int_{0}^z dz' E_z(z'),
\end{equation}
where we chose to fix the arbitrary constant $\phi(0)=0$. Next, the {\it electrostatic} potential difference between SLG and Ti is:
\begin{equation}
\label{eq:EqS6}
\phi_{SLG}-\phi_{Ti} = \frac{qn_{Ti}}{C} = \frac{q(\Delta n-n_{SLG})}{C},
\end{equation}
where $C =  \epsilon_0 \epsilon_{TiO_2}/t_{TiO_2}$ is the capacity per unit area of the junction (here $\epsilon_0$ is the vacuum permittivity, while $\epsilon_{TiO_2}\sim$15\cite{WangACSAEM2019} is the relative dielectric constant of the oxide barrier and $t_{TiO_2}$ its thickness).

We now calculate the chemical potential in SLG and Ti.
In the Ti electrode the chemical potential is equal to $\mu_{\rm Ti}=-q\phi_{Ti}-q\Phi_{Ti}$ where $\phi_{Ti}$ is the electrostatic potential in the in Ti and $q\Phi_{Ti} \approx 4.3~{\rm eV}$ the Ti work function.
In SLG the chemical potential is given instead by $\mu_{SLG}= -q\phi_{SLG}-q\Phi_{SLG}^{0}+\Delta\mu_{SLG}$, where $ \phi_{SLG}$ is the electrostatic potential in SLG, $q\Phi_{SLG}^{0}\approx 4.6~{\rm eV}$ is the work function of {\it undoped} graphene, and $\Delta\mu_{SLG} =  v_{\rm D} \text{sgn}(n_{SLG})\sqrt{\pi |n_{SLG}|}$ is the shift in chemical potential due to doping, $v_{\rm D} \approx 10^6\,{\rm  m/s}$ being the Dirac velocity in graphene and $n_{SLG}$ its electronic density. The external voltage source imposes a difference in chemical potential of $-qV_{\rm bias}$ between Ti and SLG, \textit{i.e.}:
\begin{equation}
\label{eq:EqS7}
-qV_{\rm bias} = \mu_{\rm Ti} - \mu_{SLG}.
\end{equation}
Substituting Eq.\ref{eq:EqS7} in Eq.\ref{eq:EqS6} we obtain the equilibrium equation (Eq.\ref{eq:Eq1}) in the main text. In principle Eq.\ref{eq:Eq1} should be solved self-consistently together with the Poisson equation in the CQD region.
To simulate our device we neglect, for the sake of simplicity, $\Delta n$ in the absence of light.
This approximation can be justified as the chemical potential of isolated CQD is not far from the charge neutrality point of SLG.
Moreover, simulations based on this approximation reproduce fairly well experimental data. 
Neglecting $\Delta n$, Eq.\ref{eq:Eq1} can be solved numerically for $\Delta \mu_{SLG}$ as a function of $V_{\rm bias}$.

\subsection{Conduction through the oxide barrier}
The current flowing through the junction can be calculated following for example Refs.\citenum{SuppBritS2012,SuppAliaAPL2016}.
We consider the current as generated by instantaneous energy-conserving jumps between Ti and SLG.
At a given energy $\epsilon$, the contribution to the current coming from states at energy between $\epsilon$ and $\epsilon+d\epsilon$ is:
\begin{equation}
\label{eq:EqS8}
\gamma A_{\rm ph}D_{SLG}(\epsilon)D_{Ti}(\epsilon)[f_{SLG}(\epsilon)-f_{Ti}(\epsilon)]T(\epsilon)d\epsilon,
\end{equation}
where $D_{SLG}(\epsilon)$ and $D_{Ti}(\epsilon)$ are the densities of states per unit area of SLG and Ti, which measure the number of possible initial and final states; $f_{SLG/Ti} (\epsilon)\equiv f[(\epsilon-\mu_{SLG/Ti})/(k_{\rm B}T)]$, with $f(x) =(1+e^x)^{-1}$, are the Fermi-Dirac occupation factors that measure the availability of the states for conduction, $T(\epsilon)$ is the probability for an electron injected in the barrier at energy $\epsilon$ to cross it, and $\gamma$ is a constant that takes into account the probability of transition from SLG to the oxide and from oxide to the metal (and {\it vice versa}).
We expect the density of state of Ti to be slowly varying in the energy range where transitions occur. We therefore approximate it with a constant and regroup $D_{Ti}$ and $\gamma$ in a single constant (with the units of a current) $I_0 = \gamma D_{Ti}$.
The density of states of SLG reads instead:
\begin{equation}
\label{eq:EqS9}
D_{SLG}(\epsilon) = \tilde{D}\left[\epsilon - (-q\phi_{SLG}-\Phi^0_{SLG})\right],
\end{equation}
where $\tilde{D}(\epsilon) =  2|\epsilon|/(\pi \hbar^2 v_{\rm D}^2)$ is the density of states per unit area of SLG referred to the charge neutrality (Dirac) point, and $-q\phi_{SLG}-\Phi^0_{SLG}$ is the energy of the Dirac point (See Fig.\ref{fig:FigS2}).

The conduction across the barrier happens by transmission or tunnelling through the conduction band of TiO$_2$. Hole conduction in the valence band is strongly suppressed because of the asymmetric band alignment. We approximate the transmission probability $T(\epsilon)$ as the transmission probability of a trapezoidal barrier, \textit{i.e}:
\begin{equation}
\label{eq:EqS10}
T(\epsilon) =  T_{\rm TB}(\epsilon, -q\chi_{TiO_2}-q\phi_{Ti},-q\chi_{TiO_2}-q\phi_{SLG}),
\end{equation}
where:
%
\begin{equation}
\label{eq:EqS11}
T_{\rm TB}(\epsilon,\Delta_1,\Delta_2)  =\exp\left\{\frac{2 \alpha[\theta(\Delta_2-\epsilon)(\Delta_2-\epsilon)^{3/2}-\theta(\Delta_1-\epsilon)(\Delta_1-\epsilon)^{3/2}]}{3(\Delta_1-\Delta_2)}\right\}.
\end{equation}
%
$T_{\rm TB}(\epsilon,\Delta_1,\Delta_2)$ is the transmissivity at the energy $\epsilon$ of a one dimensional trapezoidal potential barrier\cite{Sakurai1995} whose two edges are located at the energies $\Delta_1$ and $\Delta_2$, calculated within the Wentzel-Kramers-Brillouin approximation\cite{Sakurai1995}, and $\alpha = 2t_{TiO_2}\sqrt{2 m^*_{TiO_2}}/\hbar$, with $m^*_{TiO_2}$ being the conduction band mass of TiO$_2$.
Note that the energy difference between the two sides of the barrier is due to the {\it electrostatic} potential difference (Eq.\ref{eq:EqS6}). The analysis of more complicated barrier-lowering effects, like image-charge potential, is outside the scope of this work. 

Substituting Eqs.\ref{eq:EqS9}-\ref{eq:EqS10}-\ref{eq:EqS11} into Eq.\ref{eq:EqS8}, integrating over the energy and using $T_{\rm TB}(\epsilon+\delta, \Delta_1+\delta, \Delta_2+\delta)= T_{\rm TB}(\epsilon, \Delta_1, \Delta_2)$ for every $\delta$, we obtain:
%
\begin{equation}
\label{eq:EqS12}
I = I_0 A_{\rm ph} \int_{-\infty}^\infty  \tilde{D}\left(\epsilon + \Delta\mu_{SLG} - qV_{\rm bias}/2\right)\left[f\left(\frac{\epsilon -qV_{\rm bias}/2}{k_{\rm B}T}\right)-f\left(\frac{\epsilon + qV_{\rm bias}/2}{k_{\rm B}T}\right)\right]T_{\rm TB}(\epsilon, \Delta_1, \Delta_2)d\epsilon,
\end{equation}
%
where
$\Delta_1 = -q\chi_{TiO_2} + q\Phi_{Ti} -qV_{\rm bias}/2$,
and
$\Delta_2 = -q\chi_{TiO_2} + q\Phi_{SLG}^0-\Delta\mu_{SLG} + qV_{\rm bias}/2$.
We used Eq.\ref{eq:EqS12} to calculate the theoretical curve in Fig.\ref{fig:Fig3}(a).
We can separate Eq.\ref{eq:EqS12} into three contributions.
The first one is the contribution from electrons passing above the barrier, \textit{i.e.} with $\epsilon> \max(\Delta_1,\Delta_2)$, where the transmissivity $T_{\rm TB}(\epsilon, \Delta_1, \Delta_2)$ is unity.
This is called thermoionic current $I_{\rm th}$ and reads:
%
\begin{equation}
\label{eq:EqS13}
I_{\rm th}  = I_0 A_{\rm ph} \int_{\max(\Delta_1,\Delta_2)}^\infty  D_{SLG}\left(\epsilon + \Delta\mu_{SLG} - qV_{\rm bias}/2\right)\left[f\left(\frac{\epsilon -qV_{\rm bias}/2}{k_{\rm B}T}\right)-f\left(\frac{\epsilon + qV_{\rm bias}/2}{k_{\rm B}T}\right)\right]d\epsilon.
\end{equation}
On the other side, electrons at smaller energy need quantum mechanical tunnelling to cross at least part of the barrier.
The tunnelling current is given by:
\begin{equation}
\label{eq:EqS14}
I_{\rm tunn.}   = I_0 A_{\rm ph} \int_{-\infty}^{\max(\Delta_1,\Delta_2)} D_{SLG}\left(\epsilon + \Delta\mu_{SLG} - qV_{\rm bias}/2\right)\left[f\left(\frac{\epsilon -qV_{\rm bias}/2}{k_{\rm B}T}\right)-f\left(\frac{\epsilon + qV_{\rm bias}/2}{k_{\rm B}T}\right)\right]T(\epsilon, \Delta_1, \Delta_2)d\epsilon.
\end{equation}
Moreover it is useful to separate the $k_{\rm B}T\to 0$ limit of Eq.\ref{eq:EqS14} that we dub 'standard tunnelling' current:
\begin{equation}
\label{eq:EqS15}
I_{\rm n.t.}  = I_0 A_{\rm ph} \int_{-qV_{\rm bias}/2}^{qV_{\rm bias}/2} D_{SLG}\left(\epsilon + \Delta\mu_{ SLG} - qV_{\rm bias}/2 \right)T(\epsilon, \Delta_1, \Delta_2)d\epsilon.
\end{equation}
%
The remaining part $I_{\rm a. t.} =I_{\rm tunn.} -I_{\rm n.t}$ is instead referred to as 'thermally-activated tunnelling' current.
Finally,
\begin{equation}
\label{eq:EqS16}
I = I_{\rm th} + I_{\rm a.t.} + I_{\rm n.t.}.
\end{equation}
Our calculations suggest that in this type of device the dominant contribution comes either from thermally-activated tunnelling or from thermionic emission as already suggested in Ref.\citenum{SuppShayN2017}. It  is difficult to distinguish experimentally between these two effects since they have similar temperature and barrier height dependence. Nevertheless, a distinction between these two effects is not required for modelling the photocurrent. The inclusion of more complicated barrier-lowering effects can however modify this scenario.

\subsection{Photocurrent generation}
In this section we analyse theoretically the mechanism of photocurrent generation in our photodetectors.
We thus calculate the current flowing in the device in presence of a finite incident light power.
We can consider Eq.\ref{eq:EqS12} as a function of the bias voltage $V_{\rm bias}$ and of the SLG chemical potential shift $\Delta\mu_{SLG}$. All the other parameters appearing in the integral are constant. Mathematically we have:
\begin{equation}
\label{eq:EqS17}
I=I[V_{\rm bias}, \Delta\mu_{SLG}(V_{\rm bias},P_D)],
\end{equation}
where we stress that $V_{\rm bias}$ does not depend on the illumination condition, while $\Delta\mu_{SLG}$ depends both on the bias voltage and on the incident power density.
The photocurrent is obtained as difference in the presence of light and in the dark. This yields:
%
\begin{equation}
\label{eq:EqS18}
\begin{split}
I_{\rm ph}(V_{\rm bias},P_D)& = I[V_{\rm bias}, \Delta\mu_{SLG}(V_{\rm bias},P_D)]-I[V_{\rm bias}, \Delta\mu_{SLG}(V_{\rm bias},P_D =0)]\\
&\approx \frac{\partial I[V_{\rm bias}, \Delta\mu_{SLG}(V_{\rm bias},P_D=0)]}{\partial P_D}P_D\\
& = \frac{\partial I[V_{\rm bias}, \Delta\mu_{SLG}(V_{\rm bias},P_D=0)]}{\partial \Delta\mu_{SLG}}\frac{\partial \Delta\mu_{SLG}(V_{\rm bias},P_D=0) }{\partial P_D}P_D\\
& = \frac{1}{qR_{\rm tr}}\frac{\partial \Delta\mu_{SLG}(V_{\rm bias},P_D=0) }{\partial P_D}P_D,
\end{split}
\end{equation}
%
where in the second line we approximated to the linear order in the power density (we are interested here in the low-power linear response), while in the third line we applied chain rule for derivation. In the last line we defined the transfer resistivity as:
\begin{equation}
\label{eq:EqS19}
\frac{1}{R_{\rm tr}} = q \frac{\partial I[V_{\rm bias},  \Delta\mu_{SLG}(V_{\rm bias},P_D=0)]}{\partial \Delta\mu_{SLG}}.
\end{equation}
This quantity measures how much a change in the barrier height affects the current. Note that despite having units of resistance $R_{\rm tr}$ differs from a resistance for two reasons: first, it is a differential quantity, relating current variations to voltage variations, and, second, it relates the current flowing in the device to a voltage that is not the total bias voltage. 
The transfer resistance $R_{\rm tr}$ can be calculated by differentiating Eq.\ref{eq:EqS12} under the integral sign and then performing the integral numerically. As stated in the main text $R_{\rm tr}$ can be approximated using the differential resistance in dark. We checked numerically the validity of this approximation without giving it a formal justification.

To calculate the derivative appearing in the last line of Eq.\ref{eq:EqS18} we substitute Eq.\ref{eq:Eq2} into Eq.\ref{eq:Eq1} and differentiate with respect to the power density $P_D$. We then obtain:
\begin{equation}
\label{eq:EqS20}
0 = \frac{q}{C}\left(\frac{dn_{SLG}}{d \Delta\mu_{SLG}}\frac{\partial \Delta\mu_{SLG} }{\partial P_D} - \frac{\tau_{\rm tr}{\rm QE}}{E_{\rm ph}}\right) + \frac{\partial \Delta\mu_{SLG} }{\partial P_D}.
\end{equation}
Solving for $\partial \Delta\mu_{SLG} /\partial P_D$ yields:
\begin{equation}
\label{eq:EqS21}
\frac{\partial \Delta\mu_{SLG} }{\partial P_D} = \frac{q^2 \tau_{\rm tr} {\rm QE}}{E_{\rm ph}\left(C+q^2\frac{dn_{SLG}}{d \Delta\mu_{SLG}}\right)}=\frac{q^2 \tau_{\rm tr} {\rm QE}A_{\rm ph}}{E_{\rm ph}A_{\rm ph}\left(C+C_{\rm diff}\right)}= \frac{q^2 \tau_{\rm tr} {\rm QE}A_{\rm ph}}{E_{\rm ph}C_{\rm tot}},
\end{equation}
where $C_{\rm diff} = q^2 dn_{SLG}/d \Delta\mu_{SLG}=2q^2/(\hbar v_{\rm D})\sqrt{|n_{SLG}|/\pi}$ is the differential quantum capacitance of SLG.
The total capacitance $C_{\rm tot} = A_{\rm ph}(C+C_{\rm diff})$ measures how much of the transferred charge is needed to lower the barrier at a certain voltage.

Substituting Eq.\ref{eq:EqS21} into Eq.\ref{eq:EqS18} and recalling that: 
\begin{equation}
\label{eq:EqS22}
R_{\rm ph} = \frac{I_{\rm ph}}{A_{\rm ph} P_D},
\end{equation}
we obtain Eq.\ref{eq:Eq4} of the main text.

\end{NoHyper}


\begin{thebibliography}{100}

\bibitem{Reed2004} G. T. Reed and A. P. Knights, \emph{Silicon photonics: an introduction}, (John Wiley \& Sons, 2004).

\bibitem{Sze2006} S. M. Sze and K. K. Ng, \emph{Physics of semiconductor devices}, (John Wiley \& Sons, Hoboken, NJ, 2006).

\bibitem{KangNP2009} Y. M. Kang, H. D. Liu, M. Morse, M. J. Paniccia, M. Zadka, S. Litski, G. Sarid, A. Pauchard, Y. H. Kuo, H. W. Chen, W. S. Zaoui, J. E. Bowers, A. Beling, D. C. McIntosh, X. G. Zheng and J. C. Campbell, Nat. Photonics \textbf {3}, 59 (2009).

\bibitem{WangS2011} J. A. Wang and S. Lee, Sensors \textbf {11}, 696 (2011).

\bibitem{NovoS2004} K. S. Novoselov, A. K. Geim, S. V. Morozov, D. Jiang, Y. Zhang, S. V. Dubonos, I. V. Grigorieva and A. A. Firsov, Science \textbf {306}, 666 (2004).

\bibitem{NairS2008} R. R. Nair, P. Blake, A. N. Grigorenko, K. S. Novoselov, T. J. Booth, T. Stauber, N. M. R. Peres and A. K. Geim, Science \textbf {320}, 1308 (2008).

\bibitem{MayoNL2011} A. S. Mayorov, R. V. Gorbachev, S. V. Morozov, L. Britnell, R. Jalil, L. A. Ponomarenko, P. Blake, K. S. Novoselov, K. Watanabe, T. Taniguchi and A. K. Geim, Nano Lett. \textbf {11}, 2396 (2011).

\bibitem{AkinN2019} D. Akinwande, C. Huyghebaert, C.-H. Wang, M. I. Serna, S. Goossens, L.-J. Li, H. S. P. Wong and F. H. L. Koppens, Nature \textbf {573}, 507 (2019).

\bibitem{NeumNM2019} D. Neumaier, S. Pindl and M. C. Lemme, Nat. Mater. \textbf {18}, 525 (2019).

\bibitem{GossNP2017} S. Goossens, G. Navickaite, C. Monasterio, S. Gupta, J. J. Piqueras, R. Pérez, G. Burwell, I. Nikitskiy, T. Lasanta, T. Galán, E. Puma, A. Centeno, A. Pesquera, A. Zurutuza, G. Konstantatos and F. Koppens, Nat. Photonics \textbf {11}, 366 (2017).

\bibitem{GoykNL2016} I. Goykhman, U. Sassi, B. Desiatov, N. Mazurski, S. Milana, D. De Fazio, A. Eiden, J. Khurgin, J. Shappir, U. Levy and A. C. Ferrari, Nano Lett. \textbf {16}, 3005 (2016).

\bibitem{LiuN2011} M. Liu, X. B. Yin, E. Ulin-Avila, B. S. Geng, T. Zentgraf, L. Ju, F. Wang and X. Zhang, Nature \textbf {474}, 64 (2011).

\bibitem{SoriNP2018} V. Sorianello, M. Midrio, G. Contestabile, I. Asselberghs, J. Van Campenhout, C. Huyghebaert, I. Goykhman, A. K. Ott, A. C. Ferrari and M. Romagnoli, Nat. Photonics \textbf {12}, 40 (2018).

\bibitem{KoppNN2014} F. H. L. Koppens, T. Mueller, P. Avouris, A. C. Ferrari, M. S. Vitiello and M. Polini, Nat. Nanotechnol. \textbf {9}, 780 (2014).

\bibitem{KonsNN2012} G. Konstantatos, M. Badioli, L. Gaudreau, J. Osmond, M. Bernechea, F. P. G. de Arquer, F. Gatti and F. H. L. Koppens, Nat. Nanotechnol. \textbf {7}, 363 (2012).

\bibitem{ZhanSR2014} W. J. Zhang, C. P. Chuu, J. K. Huang, C. H. Chen, M. L. Tsai, Y. H. Chang, C. T. Liang, Y. Z. Chen, Y. L. Chueh, J. H. He, M. Y. Chou and L. J. Li, Sci. Rep. \textbf {4}, 3826 (2014).

\bibitem{DeFaACS2016} D. De Fazio, I. Goykhman, D. Yoon, M. Bruna, A. Eiden, S. Milana, U. Sassi, M. Barbone, D. Dumcenco, K. Marinov, A. Kis and A. C. Ferrari, ACS Nano \textbf {10}, 8252 (2016).

\bibitem{XiaNN2009} F. N. Xia, T. Mueller, Y. M. Lin, A. Valdes-Garcia and P. Avouris, Nat. Nanotechnol. \textbf {4}, 839 (2009).

\bibitem{DeRoOE2011} C. T. DeRose, D. C. Trotter, W. A. Zortman, A. L. Starbuck, M. Fisher, M. R. Watts and P. S. Davids, Opt. Express \textbf {19}, 24897 (2011).

\bibitem{ColaIEEE2007} L. Colace, P. Ferrara, G. Assanto, D. Fulgoni and L. Nash, IEEE Photon. Technol. Lett. \textbf {19}, 1813 (2007).

\bibitem{KufeAM2015} D. Kufer, I. Nikitskiy, T. Lasanta, G. Navickaite, F. H. L. Koppens and G. Konstantatos, Adv. Mater. \textbf {27}, 176 (2015).

\bibitem{ShayN2017} M. Shaygan, Z. Wang, M. S. Elsayed, M. Otto, G. Iannaccone, A. H. Ghareeb, G. Fiori, R. Negra and D. Neumaier, Nanoscale \textbf {9}, 11944 (2017).

\bibitem{WangACSAEM2019} Z. Wang, B. Uzlu, M. Shaygan, M. Otto, M. Ribeiro, E. G. Marín, G. Iannaccone, G. Fiori, M. S. Elsayed, R. Negra and D. Neumaier, ACS Appl. Electron. Mater. \textbf {1}, 945 (2019).

\bibitem{CastRMP2009} A. H. Castro Neto, F. Guinea, N. M. R. Peres, K. S. Novoselov and A. K. Geim, Rev. Mod. Phys. \textbf {81}, 109 (2009).

\bibitem{LiS2009} X. S. Li, W. W. Cai, J. H. An, S. Kim, J. Nah, D. X. Yang, R. Piner, A. Velamakanni, I. Jung, E. Tutuc, S. K. Banerjee, L. Colombo and R. S. Ruoff, Science \textbf {324}, 1312 (2009).

\bibitem{KimN2009} K. S. Kim, Y. Zhao, H. Jang, S. Y. Lee, J. M. Kim, K. S. Kim, J.-H. Ahn, P. Kim, J.-Y. Choi and B. H. Hong, Nature \textbf {457}, 706 (2009).

\bibitem{ShayAP2017} M. Shaygan, M. Otto, A. A. Sagade, C. A. Chavarin, G. Bacher, W. Mertin and D. Neumaier, Ann. Phys. \textbf {529}, 1600410 (2017).

\bibitem{DasNN2008} A. Das, S. Pisana, B. Chakraborty, S. Piscanec, S. K. Saha, U. V. Waghmare, K. S. Novoselov, H. R. Krishnamurthy, A. K. Geim, A. C. Ferrari and A. K. Sood, Nat. Nanotechnol. \textbf {3}, 210 (2008).

\bibitem{GeimNM2007} A. K. Geim and K. S. Novoselov, Nat. Mater. \textbf {6}, 183 (2007).

\bibitem{XiaNN2009-2} J. Xia, F. Chen, J. Li and N. Tao, Nat. Nanotechnol. \textbf {4}, 505 (2009).

\bibitem{BritS2012} L. Britnell, R. V. Gorbachev, R. Jalil, B. D. Belle, F. Schedin, A. Mishchenko, T. Georgiou, M. I. Katsnelson, L. Eaves, S. V. Morozov, N. M. R. Peres, J. Leist, A. K. Geim, K. S. Novoselov and L. A. Ponomarenko, Science \textbf {335}, 947 (2012).

\bibitem{AliaAPL2016} I. Aliaj, I. Torre, V. Miseikis, E. di Gennaro, A. Sambri, A. Gamucci, C. Coletti, F. Beltram, F. M. Granozio, M. Polini, V. Pellegrini and S. Roddaro, APL Mater. \textbf {4}, 066101 (2016).

\bibitem{Griffiths2018} D. J. Griffiths and D. F. Schroeter, \emph{Introduction to quantum mechanics}, (Cambridge University Press, 2018).

\bibitem{MichNP2010} J. Michel, J. F. Liu and L. C. Kimerling, Nat. Photonics \textbf {4}, 527 (2010).

\bibitem{KaniOER2004} J. Kaniewski and J. Piotrowski, Opto-Electron. Rev. \textbf {12}, 139 (2004).

\bibitem{Brown1992} R. G. Brown and P. Y. Hwang, \emph{Introduction to random signals and applied Kalman filtering}, (Wiley New York, 1992).

\bibitem{BalaNN2013} A. A. Balandin, Nat. Nanotechnol. \textbf {8}, 549 (2013).

\bibitem{ZhanACSN2011} Y. Zhang, E. E. Mendez and X. Du, ACS Nano \textbf {5}, 8124 (2011).

\bibitem{KarnNC2016} P. Karnatak, T. P. Sai, S. Goswami, S. Ghatak, S. Kaushal and A. Ghosh, Nat. Commun. \textbf {7}, 13703 (2016).

\bibitem{MavrN2018} N. Mavredakis, R. Garcia Cortadella, A. Bonaccini Calia, J. A. Garrido and D. Jiménez, Nanoscale \textbf {10}, 14947 (2018).

\bibitem{HoogRPP1981} F. N. Hooge, T. G. M. Kleinpenning and L. K. J. Vandamme, Rep. Prog. Phys. \textbf {44}, 479 (1981).

\bibitem{HoogIEEE1994} F. N. Hooge, IEEE Trans. Electron Dev. \textbf {41}, 1926 (1994).

\bibitem{ZhigJCTE2007} G. P. Zhigal’skii, A. A. Gvas’kov and P. O. Sitkin, J. Commun. Technol. Electron. \textbf {52}, 701 (2007).

\bibitem{MonrIEEE2000} E. Monroy, F. Calle, J. L. Pau, E. Munoz and F. Omnes, Electron. Lett. \textbf {36}, 2096 (2000).

\bibitem{StolAPL2015} M. A. Stolyarov, G. Liu, S. L. Rumyantsev, M. Shur and A. A. Balandin, Appl. Phys. Lett. \textbf {107}, 023106 (2015).

\end{thebibliography}

\begin{thebibliography}{20}

\bibitem{LeeS2016} J. W. Lee, D. Y. Kim, S. Baek, H. Yu and F. So, Small \textbf {12}, 1328 (2016).

\bibitem{MihiAM2014} A. Mihi, F. J. Beck, T. Lasanta, A. K. Rath and G. Konstantatos, Adv. Mater. \textbf {26}, 443 (2014).

\bibitem{IpNN2012} A. H. Ip, S. M. Thon, S. Hoogland, O. Voznyy, D. Zhitomirsky, R. Debnath, L. Levina, L. R. Rollny, G. H. Carey, A. Fischer, K. W. Kemp, I. J. Kramer, Z. Ning, A. J. Labelle, K. W. Chou, A. Amassian and E. H. Sargent, Nat. Nanotechnol. \textbf {7}, 577 (2012).

\bibitem{SuppKonsNN2012} G. Konstantatos, M. Badioli, L. Gaudreau, J. Osmond, M. Bernechea, F. P. G. de Arquer, F. Gatti and F. H. L. Koppens, Nat. Nanotechnol. \textbf {7}, 363 (2012).

\bibitem{FerrPRL2006} A. C. Ferrari, J. C. Meyer, V. Scardaci, C. Casiraghi, M. Lazzeri, F. Mauri, S. Piscanec, D. Jiang, K. S. Novoselov, S. Roth and A. K. Geim, Phys. Rev. Lett. \textbf {97}, 187401 (2006).

\bibitem{FerrNN2013} A. C. Ferrari and D. M. Basko, Nat. Nanotechnol. \textbf {8}, 235 (2013).

\bibitem{CancNL2011} L. G. Cancado, A. Jorio, E. H. Ferreira, F. Stavale, C. A. Achete, R. B. Capaz, M. V. Moutinho, A. Lombardo, T. S. Kulmala and A. C. Ferrari, Nano Lett. \textbf {11}, 3190 (2011).

\bibitem{BiN2019} Y. Bi, A. Bertran, S. Gupta, I. Ramiro, S. Pradhan, S. Christodoulou, S.-N. Majji, M. Z. Akgul and G. Konstantatos, Nanoscale \textbf {11}, 838 (2019).

\bibitem{YuNL2009} Y.-J. Yu, Y. Zhao, S. Ryu, L. E. Brus, K. S. Kim and P. Kim, Nano Lett. \textbf {9}, 3430 (2009).

\bibitem{SeoJAP2014} J.-T. Seo, J. Bong, J. Cha, T. Lim, J. Son, S. H. Park, J. Hwang, S. Hong and S. Ju, J. Appl. Phys. \textbf {116}, 084312 (2014).

\bibitem{BoroPCCP2011} A. Borodin and M. Reichling, Phys. Chem. Chem. Phys. \textbf {13}, 15442 (2011).

\bibitem{SuppWangACSAEM2019} Z. Wang, B. Uzlu, M. Shaygan, M. Otto, M. Ribeiro, E. G. Marín, G. Iannaccone, G. Fiori, M. S. Elsayed, R. Negra and D. Neumaier, ACS Appl. Electron. Mater. \textbf {1}, 945 (2019).

\bibitem{Weast1988} R. C. Weast, M. J. Astle and W. H. Beyer, \emph{CRC handbook of chemistry and physics}, (CRC press Boca Raton, FL, 1988).

\bibitem{SuppBritS2012} L. Britnell, R. V. Gorbachev, R. Jalil, B. D. Belle, F. Schedin, A. Mishchenko, T. Georgiou, M. I. Katsnelson, L. Eaves, S. V. Morozov, N. M. R. Peres, J. Leist, A. K. Geim, K. S. Novoselov and L. A. Ponomarenko, Science \textbf {335}, 947 (2012).

\bibitem{SuppAliaAPL2016} I. Aliaj, I. Torre, V. Miseikis, E. di Gennaro, A. Sambri, A. Gamucci, C. Coletti, F. Beltram, F. M. Granozio, M. Polini, V. Pellegrini and S. Roddaro, APL Mater. \textbf {4}, 066101 (2016).

\bibitem{Sakurai1995} J. J. Sakurai and E. D. Commins, \emph{Modern quantum mechanics}, (American Association of Physics Teachers, 1995).

\bibitem{SuppShayN2017} M. Shaygan, Z. Wang, M. S. Elsayed, M. Otto, G. Iannaccone, A. H. Ghareeb, G. Fiori, R. Negra and D. Neumaier, Nanoscale \textbf {9}, 11944 (2017).


\end{thebibliography}
\end{document}